\documentstyle[12pt]{article}
\begin{document}
\begin{titlepage}

\centerline
{\bf{ Cutoff and lattice effects in the $\varphi^4$ theory 
of confined systems }}

\vspace*{1cm}
\centerline {X. S. Chen $^{1,2}$ and V. Dohm $^{1}$}
\vspace{0.8cm} 
\centerline{$^1$Institut f\"ur Theoretische Physik, Technische Hochschule 
Aachen,}
\centerline{D-52056 Aachen, Germany}     
\vspace{0.8cm}
\centerline{$^{2}$ Institute of Particle Physics, Hua-Zhong Normal University,}

\centerline{Wuhan 430079, China}
 
\vspace{2cm}

\begin{abstract}
We study cutoff and lattice effects in the $O(n)$ symmetric
$\varphi^4$ theory for a $d$-dimensional cubic geometry of size
$L$ with periodic boundary conditions. In the large-$n$ limit above
$T_c$, we show that $\varphi^4$ field theory at finite cutoff  
$\Lambda$ predicts the  nonuniversal deviation 
$\sim (\Lambda L)^{-2}$ from asymptotic bulk critical behavior 
that violates finite-size scaling and disagrees with the 
deviation $\sim e^{-cL}$ that we find in the $\varphi^4$ lattice model. 
The exponential size dependence requires a non-perturbative treatment
of the $\varphi^4$ model. Our arguments indicate that these results should 
be valid for general $n$ and $d > 2$.
\\
\end{abstract}

\vspace{1.5cm}

\noindent
PACS numbers: 05.70.Jk, 64.60.i, 75.40.Mg
\end{titlepage}

\newpage
The concept of finite-size scaling   
\cite{F} plays a fundamental role in the theory of finite-size effects
near phase transitions [1-4] and is indispensable for the analysis of
numerical studies of critical phenomena in small systems \cite{Bi1}.
Consider, for example, the susceptibility $\chi(T,L)$ of a ferromagnetic
system for $T \geq T_c$ in a finite geometry of size $L$. Finite-size
scaling is expected to be valid for large $L$ and large correlation
length $\xi \sim (T - T_c)^{-\nu}$, with a scaling form 
$ \chi(T,L) \; = \; L^{\gamma/\nu} f(L/\xi) $
where $\gamma$ and $\nu$ are bulk critical exponents and where the scaling
function $f$ depends on the geometry and boundary conditions but not on any 
other length scale. In this paper we shall consider only periodic boundary 
conditions and cubic geometry, $V = L^d$.

Finite-size scaling functions have been calculated within the $O(n)$ symmetric
$\varphi^4$ field theory in $2 < d < 4$ dimensions [6-9] and quantitative 
agreement with Monte-Carlo (MC)
data has been found [8-10]. It is the
purpose of the present Rapid Note to call attention to a remarkable feature
that has not been explained by the field-theoretic calculations. This is the
exponential (rather than power-law) approach
\begin{equation}
\Delta \chi \; \equiv \; \chi(T, \infty) \; - \; \chi(T,L) \sim \exp 
[-\Gamma(T)L]
\end{equation}
towards the asymptotic bulk critical behavior 
$\chi(T, \infty) \sim \xi^{\gamma/\nu}$
above $T_c$, as has been found in several exactly solvable model systems 
[1,2,11-14]. By contrast, field theory [6-9]
implies a non-exponential behavior $\Delta \chi \sim O ((L/\xi)^{-d})$ in
one-loop order above $T_c$ for $d < 4$. 
We are not aware of numerical tests of this property, e.g., 
by MC simulations.

We shall analyze this problem on the basis of the exact result \cite{CD1}
for $\chi$ in the large-$n$ limit of the $\varphi^4$ model. In particular 
we shall study the effect of a finite
cutoff $\Lambda$ and of a finite lattice spacing in the field-theoretic
and lattice version of the $\varphi^4$ theory. We find that field theory
at finite cutoff predicts the leading nonuniversal deviation 
$ \Delta \chi \sim  (\Lambda L)^{-2}$ from bulk critical behavior 
that violates finite-size
scaling for $d > 2$ and differs from Eq. (1). 
This is in contrast to the general belief [3-10, 15-23] (and corrects our
recent statement \cite{CD1,CD3}) that the finite-size scaling functions of the
$\varphi^4$ field theory  are universal for $2 < d < 4$ (for cubic geometry 
and periodic boundary conditions). We shall show  
that the $\varphi^4$  {\it{lattice}} theory with a finite lattice 
spacing accounts for the 
exponential size-dependence of Eq. (1).
We shall argue that a loop expansion destroys this exponential form and
that a non-perturbative  treatment \cite{CDS} of the $\varphi^4$ theory 
is required.

The $\varphi^4$ field theory is based on the statistical weight $\exp(-H)$ with
the Landau-Ginzburg-Wilson continuum Hamiltonian
\begin{equation}
H \; = \; \int_V d^dx \; \left[ {r_0\over 2} \varphi^2_0 \; + \; \frac{1}{2}
(\bigtriangledown \varphi_0)^2 \; + \; u_0(\varphi^2_0)^2  \right] \; ,
\end{equation}
with $
r_0 \; = \; r_{0c} \; + \; a_0 t \; , \; t = (T - T_c)/T_c
$ where the $n$-component field $\varphi_0(x)$ has spatial variations on
length scales larger than a microscopic length $\tilde{a}$ corresponding
to a finite cutoff $\Lambda = \pi/\tilde{a}$. Since we wish to perform
a convincing comparison with the finite-size effects of lattice systems
which have a finite lattice constant $\tilde{a}$ we must keep 
$\Lambda$ finite even if a well defined limit $\Lambda \rightarrow \infty$
can formally be performed at fixed $r_0 - r_{0c}$ for $2 < d < 4$ [6,16,17].
It is well known that this limit is justified for bulk systems \cite{ZJ}
where finite-cutoff effects are only subleading corrections to the leading
bulk critical temperature dependence.

Here we raise the question what kind
of finite-size effects exist at finite $\Lambda$. This question was left
unanswered in the renormalization-group arguments of Br\'{e}zin \cite{B} and
in the explicit field-theoretic calculations of Refs. [6-10, 17-22] which were
performed only in the limit $\Lambda \rightarrow \infty$ and where it was
tacitly assumed that finite-cutoff effects are negligible for $d < 4$. We shall
prove for $d > 2$ and $n \rightarrow \infty$ that this assumption is 
not generally justified for the field-theoretic $\varphi^4$ model for finite
systems.

We shall first examine the susceptibility of the field-theoretic model
\begin{equation}
\chi \, = \; (1/n) \int_V d^dx < \varphi_0(x)\varphi_0(0)>
\end{equation}
in the large-$n$ limit at fixed $u_0n$. For cubic geometry, $V = L^d$, the
exact result for $d > 2$ is determined by the implicit equation \cite{CD1}
\begin{equation}
\chi^{-1}  =  r_0 - r_{0c} - 4 u_0 n \tilde{\Delta}_1 
+ 4 u_0 n \left\{ \chi L^{-d} - \chi^{-1} \int\limits_{\bf{k}} 
\left[ {\bf{k^2}} (\chi^{-1} + {\bf{k^2}}) \right]^{-1}\right\},
\end{equation}
\begin{equation}
\tilde{\Delta}_1 =  \int\limits_{\bf{k}} 
(\chi^{-1} + {\bf{k}}^2)^{-1} \; - \;
L^{-d} \; \sum\limits_{{\bf{k}} \neq {\bf{0}}} (\chi^{-1} \; + \; 
{\bf{k}}^2)^{-1},
\end{equation}
where $r_{0c} = - 4u_0n \int_{\bf{k}} {\bf{k}}^{-2}$. 
Here $\int_{\bf{k}}$ stands for $(2 \pi)^{-d} \int d^dk$ with 
$|k_j| \leq  \Lambda$, and the summation $\sum_{{\bf{k}} \neq {\bf{0}}}$
runs over discrete ${\bf{k}}$ vectors with components 
$k_j = 2 \pi m_j/L, m_j = \pm 1, \pm 2, \cdots, j = 1,2,\cdots, d,$
in the range $-\Lambda \leq k_j < \Lambda$. For large $L$ at finite $\Lambda$
we have found for $d > 2$
\begin{equation}
\tilde{\Delta}_1 =  I_1(\chi^{-1}L^2) \; L^{2-d}
+\Lambda^{d-2} \left\{ a_1 (d, \chi^{-1} \Lambda^{-2}) (\Lambda L)^{-2}
+ \quad  O \left[  (\Lambda L)^{-4} \right]\right\} ,
\end{equation}
\begin{equation}
I_1(x) = - (2\pi)^{-2}  \int\limits^{\infty}_{0} \;  dy \; e^{-(xy/4\pi^2)}
\; \left[ K (y)^d - (\pi/y)^{d/2} \; - \; 1  \right] \quad , 
\end{equation}
\begin{equation}
a_1(d, \chi^{-1} \Lambda^{-2})\; = \; 
\frac{d}{3(2\pi)^{d-2}} \; \int\limits^{\infty}_{0} dx \; x 
\left[ \int^1_{-1} dy \; e^{-y^2x} \right]^{d-1}
\; \exp \left[ - (1 + \chi^{-1} \Lambda^{-2})x \right]
\end{equation}
with $ K(y)  \; = \; \sum\limits_{m = - \infty}^{\infty}\; e^{-ym^2}$.
Near $T_c$, i.e., for small $\chi^{-1} \Lambda^{-2}$ at finite $\Lambda$,
the bulk integral in Eq. (4) yields for $2 < d < 4$
\begin{equation}
\int\limits_{\bf{k}}
\left[ {\bf{k}}^2 (\chi^{-1}\;  + \; {\bf{k}}^2 ) \right] ^{-1} \; = \;
A_d \;  \chi^{\epsilon/2}\epsilon^{-1} \left\{ 1 + O 
\left[ (\chi^{-1} \Lambda^{-2})^{\epsilon/2} \right] \right\}
\end{equation}
with $\epsilon = 4 - d$ and 
$A_d \; = \; 2^{2-d} \pi^{-d/2} (d - 2)^{-1} \Gamma (3 - d/2) \quad .
$
This leads to the large-$L$ and small-$t$ representation at finite $\Lambda$
\begin{equation}
\chi \; = \; L^{\gamma/\nu} \; P(t(L/\xi_0)^{1/\nu}, \Lambda L)
\end{equation}
where the function $P$ is determined implicitly by 
\begin{eqnarray}
P^{-1/\gamma} \; = \; t(L/\xi_0)^{1/\nu} \;  +  \;
\epsilon A_d^{-1} \left[ P \; - \; I_1(P^{-1}) - 
a_1 (d, 0)(\Lambda L)^{d-4} 
\right] \; ,  
\end{eqnarray}
apart from $ O \left[ (\Lambda L)^{d-6} \right]$ corrections,
with the critical exponents $\nu = (d - 2)^{-1}$ and $\gamma = 2/(d-2)$, and 
with the bulk correlation-length amplitude $\xi_0$ \cite{CD1}.
We note that the term $I_1(P^{-1})$ is a  ${\bf{k}} \neq {\bf{0}}$ contribution
whereas the term $\sim P$ on the r.h.s. of Eq. (11) comes from the 
${\bf{k}} = {\bf{0}}$ mode.

At first sight, the $\Lambda$ dependent term in Eq. (11) seems to be a 
subleading correction and appears to be negligible for large $L$. This is 
asymptotically
correct as long as $P - I_1(P^{-1}) > 0$ does not vanish in the large-$L$ 
limit.
This is indeed the case for $t(L/\xi_0)^{1/\nu} < \infty$, i.e., as long as
the critical point is approached at finite ratio $L/\xi$ .
This corresponds to paths in the $L^{-1} - \xi^{-1}$ plane (Fig. 1) that 
approach the origin 
$L^{-1} = 0, \xi^{-1} = 0 $ along curves
with a non-vanishing asymptotic slope $\xi/L > 0$. Along these paths the 
function $P$ remains finite and hence $P - I_1(P^{-1})$ remains non-zero 
(positiv)
which was tacitly assumed previously \cite{CD1} where the $\Lambda$-dependent
terms in Eq. (11) were dropped (see Eq. (62) of Ref. \cite{CD1}).

There exist, however, significant paths in the $L^{-1} - \xi^{-1}$ plane where
$t(L/\xi_0)^{1/\nu}$ becomes arbitrarily large. This includes paths 
at constant $t > 0$ or $\xi < \infty$ with increasing $L$
corresponding to an approach towards the asymptotic bulk value $\chi_b$ 
(arrow in Fig. 1). We emphasize that these paths lie entirely in the
asymptotic region $\xi \gg \Lambda^{-1}, L \gg \Lambda^{-1}, \chi_b = \xi^2 \gg
\Lambda^{-2}$. In such limits the quantity $P \sim (\xi/L)^{\gamma/\nu}$
approaches zero. As a remarkable feature we find that in Eq. (11) the function
$I_1(P^{-1})$ (which originates from the ${\bf{k}} \neq {\bf{0}}$ modes)
completely cancels the term $P$ (which comes from the ${\bf{k}} = {\bf{0}}$ 
mode) according to the small-$P$ representation
\begin{equation}
I_1 (P^{-1}) \; = \; P \; + \; O 
\left[ P^{(3-d)/4} \exp (-P^{-1/2}) \right]\; .
\end{equation}
In other words, the higher-mode contribution $\sim I_1(P^{-1})$ does not 
represent a ''correction'' to the lowest-mode term $P$ but becomes as large as
the lowest-mode term itself. This result is quite plausible because above 
$T_c$, at fixed temperature $T - T_c > 0$, the lowest mode does not play
a significant role and does not become dangerous in the bulk limit, unlike 
the case $T \le T_c$  where the separation of the lowest mode is an important
concept \cite{B1, RGJ}.

The crucial consequence of Eq.(12) is that the term 
$P - I_1(P^{-1})$ in Eq. (11) is reduced to the exponentially small 
contribution $\sim \exp (-P^{-1/2}) \sim \exp (-L/\xi)$.
This implies that the leading finite-size deviation from bulk critical behavior
is now governed by the cutoff-dependent power-law term $(\Lambda L)^{d-4}$ in
Eq. (11) which was dropped in Ref. \cite{CD1}. This leads to the explicitly
$\Lambda$ dependent result, at finite $t \ll 1$ and finite $\Lambda L \gg 1$,
\begin{equation}
P(t(L/\xi_0)^{1/\nu}, \Lambda L) \; = \;  
\left[ t(L/\xi_0)^{1/\nu} \; 
- \; \epsilon A_d^{-1} a_1 (d,0)(\Lambda L)^{d-4} 
\right]^{-\gamma}\quad ,
\end{equation}
\begin{equation}
\chi  \; = \;  \chi_b \left[1+ \frac{\epsilon \; 2^{d-1} \pi^{d/2}} 
{\Gamma (3-d/2)}\;
a_1 (d,0)(\Lambda \xi)^{d-2}(\Lambda L)^{-2}\right],
\end{equation}
apart from $O \left[ (\Lambda L)^{-4} \; , \; e^{-L/\xi} \right]$
corrections. 
Eq. (14) is valid for $2 < d < 4$ and is applicable to the region below the
dotted line in Fig. 1. This line is a representative of a smooth 
crossover region and may be defined by requiring that the cutoff dependent
term in Eq. (11) is as large as the term $P - I_1 (P^{-1})$. In the latter
term, $P$ can be approximated by $(L/\xi)^{-\gamma/\nu}$, i.e., the
dotted line is determined  by
\begin{equation}
(L/\xi)^{-\gamma/\nu} \; - \; I_1 ((L/\xi)^{\gamma/\nu}) \; = \; a_1(d,0)
(\Lambda L)^{d-4}\quad .
\end{equation}
Eq. (15) represents a line in a crossover region separating the scaling
region (where cutoff effects can be considered as small corrections)
from the nonscaling region (where cutoff effects are dominant) close the
bulk limit.

The power law $\sim (\Lambda L)^{-2}$ in Eq. (14) disagrees with the
exact result for the spherical model on a lattice where an exponential $L$
dependence, analogous to Eq. (1), has been found \cite{BF, SR} for general
$d > 2$. This proves that $\varphi^4$ field theory at finite cutoff does not
correctly describe the leading finite-size deviations from bulk critical 
behavior of spin systems  on a lattice above $T_c$, not only for $d > 4$, as
stated in Ref. \cite{CD3}, but more generally for $d > 2$, at least in the
large-$n$ limit. Furthermore, the result in Eq. (14) violates 
finite-size scaling {\it{in the asymptotic region}} where 
$L^{-\gamma/\nu} \chi$ should only depend on $L/\xi$, not on $\Lambda L$.
Thus $\varphi^4$ field theory at finite cutoff is inconsistent with usual
finite-size scaling not only for $d > 4$, but more generally for $d > 2$,
at least in the large-$n$ limit. This is not in conflict with the 
renormalization-group arguments of Br\'{e}zin \cite{B} who considered
only the limit of infinite cutoff in which the non-scaling region (Fig. 1)
shrinks to zero. The existence of the non-scaling region for the 
field-theoretic $\varphi^4$ model below four dimensions has been overlooked
in Sect. 4.1 of our recent work \cite{CD1}.

In the following we   briefly analyze the corresponding properties 
in the $\varphi^4$ lattice model for $d > 2$.
The $\varphi^4$ lattice Hamiltonian reads
\begin{equation}
\hat{H} (\varphi_i) = \tilde{a}^d 
\left\{ \sum\limits_i \left[ \frac{\hat{r}_0}{2}\varphi^2_i 
+ \hat{u}_0 (\varphi^2_i)^2 \right] \; + \; \sum\limits_{ij}
\frac{1}{2 \tilde{a}^2} J_{ij} (\varphi_i - \varphi_j)^2 \right \}
\end{equation}
where $\tilde{a}$ is the lattice constant. 
As noted recently \cite{CD1}, the susceptibility $\hat{\chi}$ of  the 
lattice model is 
obtained from $\chi$ of the field-theoretic  model by the replacement
${\bf{k}}^2 \rightarrow \hat{J}_{\bf{k}}$ in the sums and integrals in 
Eqs. (4) and (5), where
\begin{equation}
\hat{J}_{\bf{k}} \; = \; \frac{2}{\tilde{a}^2} \left[ J(0) \; - \; J(\bf{k}) 
\right]  \; = \; J_0{\bf{k}}^2  \; + \; O(k^2_i \; k^2_j)\quad ,
\end{equation}
\begin{equation}
J({\bf{k}}) \; = \; (\tilde{a}/L)^d \; \sum\limits_{ij} 
\; J_{ij} \; e^{i{\bf{k}}\cdot ({\bf{x}}_i-{\bf{x}}_j)} \; ,
\end{equation} 
\begin{equation}
J_0 \; = \; \frac{1}{d} (\tilde{a}/L)^d  \; \sum\limits_{ij} 
(J_{ij}/\tilde{a}^2)({\bf{x}}_i - {\bf{x}}_j)^2 \quad .
\end{equation}
The crucial difference between the field-theoretic and lattice versions of the
$\varphi^4$ model comes from the large-$L$ behavior of the lattice version of
the quantity $\tilde{\Delta}_1$ in Eq. (5). Instead of Eq. (6) we now obtain
for $L \gg \tilde{a}$
\begin{equation}
\int\limits_{\bf{k}}(\hat{\chi}^{-1} \; + \;  \hat{J}_{\bf{k}})^{-1} \; - \; 
L^{-d} \sum\limits_{{\bf{k}} \neq {\bf{0}}} (\hat{\chi}^{-1} \; + \; 
\hat{J}_{\bf{k}})^{-1} = 
J_0^{-1} \;  I_1  (J^{-1}_0 \hat{\chi}^{-1} L^2) L^{2-d},
\end{equation}
apart from more rapidly vanishing terms. We have found that such terms
are only exponential (rather than power-law) corrections in the regime 
$L \gg \xi$.  This implies that, for the lattice model
in the regime $L \gg \xi$, Eq. (11) is reduced to
\begin{equation}
\hat{P}^{-1/\gamma} \; = \; t(L/\hat{\xi}_0)^{1/\nu} \quad + \quad 
\epsilon A^{-1}_d
\left[\hat{P} \; - \; I_1 (\hat{P}^{-1}) \right]
\end{equation}
without power-law corrections. This corresponds  to Eq. (77) 
of Ref. \cite{CD1}.
Here $\hat{\xi}_0$ is the bulk correlation-length amplitude of 
the lattice model and
$\hat{P} \; = \; \hat{\chi} L^{-\gamma/\nu} \; J_0 $ \cite{CD1}.
Because of the exponential behavior of $\hat{P} - I_1 (\hat{P}^{-1})$
according to Eq. (12) and because of the exponential corrections to Eq. (20) 
we see that the lattice $\varphi^4$ model indeed predicts
an exponential size dependence for $\Delta \hat{\chi}$. 
The detailed form of the ($L$-dependent) amplitude of this exponential 
size-dependence is nontrivial and will be analyzed elsewhere.

In the following we extend our analysis to the case $n = 1$ of the
field-theoretic model for  $2 < d < 4$.
The bare perturbative expressions for the effective parameters given in Eqs.
(68) - (71) of Ref. \cite{CD4} for the field-theoretic $\varphi^4$
model are valid for general $d > 2$. Application to the critical region for 
$2 < d < 4$ requires to renormalize these expressions by the 
$L$-independent $Z$-factors of the
{\it{bulk}} theory. We recall that the bulk renormalizations can well 
be performed at finite $\Lambda$ \cite{BLZ}.
This does not eliminate the cutoff dependent term 
$\sim (\Lambda L)^{-2}$ in $r^{eff}_0$
for the field-theoretic model and implies that $\Delta \chi$ will exhibit the
leading size dependence $\sim (\Lambda L)^{-2}$ above $T_c$ also for $n = 1$,
$2 < d < 4$.

A grave consequence of these results is that universal finite-size
scaling near the critical point of a finite system with periodic boundary
conditions is less generally valid than believed previously [1-24]. 
Finite-size scaling is not valid in the region below the dotted line of the 
$L^{-1} - \xi^{-1}$ plane (Fig. 1), at least in the large-$n$ limit, for the
field-theoretic $\varphi^4$ model at finite cutoff for $d > 2$. This region is
of significant interest as it describes the leading finite-size deviations from
asymptotic bulk critical behavior. The violation of finite-size scaling in 
this region originates from the $(\bigtriangledown \varphi)^2$
term in Eq. (2) that approximates  the interaction term 
$J_{ij}(\varphi_i - \varphi_j)^2$ of the $\varphi^4$ lattice Hamiltonian,
Eq. (16). 
The serious defect of this approximation at finite $\Lambda$ becomes more and 
more significant as $L/\xi \gg 1$
increases (arrow in Fig. 1) whereas it is negligible for $\xi/L > 1$. 
This defect does not show up in the $\Lambda \rightarrow  \infty$ version (or
dimensionally regularized version) of renormalized field theory.
For a discussion of the case $d > 4$
we refer to \cite{CD5}.

From the one-loop finite-size scaling functions of Ref. \cite{EDC} we find
the nonexponential behavior $\Delta \chi \sim O  ((L/\xi)^{-d})$ 
for $n = 1$ and  $2 < d < 4$ 
above $T_c$. 
The same behavior exists already in the lowest-mode approximation. 
The question arises whether higher-loop calculations would change
this $L$ dependence.  The corresponding question for 
$n \rightarrow \infty$ can be answered on the basis of our exact solution 
for $\hat{\chi}$ of the $\varphi^4$ lattice model \cite{CD1}. 
Approximating this solution
by a one-loop type expansion around the lowest-mode structure leads
to the large-$L$ behavior $\Delta \hat{\chi} \sim O ((L/\xi)^{-d})$
above $T_c$ rather than   $\sim e^{-cL}$. Thus, at least for 
$n \rightarrow \infty$,
the exponential size dependence  is a {\it{non-perturbative}}
feature. We expect, therefore, that a conclusive answer of our question
requires a non-perturbative treatment of the $\varphi^4$ lattice theory.
Our previous nonperturbative order-parameter distribution function
\cite{CDS} is an appropriate basis for analyzing  this problem which
will presumably lead to an exponential size dependence for $\Delta \hat{\chi}$ 
within the $\varphi^4$ lattice model for general $n$ above two dimensions.

It would be interesting to test the leading finite-size deviations from bulk
critical behavior by Monte-Carlo simulations. 
The absence of terms $\sim (\Lambda L)^{-2}$ would provide  
evidence for the failure of the continuum approximation 
$\sim (\bigtriangledown \varphi)^2$ of the $\varphi^4$ field theory at finite
$\Lambda$ for confined lattice
systems with periodic boundary conditions. \\

{\bf{Acknowledgment}}\\

Support by Sonderforschungsbereich 341 der Deutschen Forschungsgemeinschaft
and by NASA under contract numbers 960838 and 100G7E094 is acknowledged. 
One of the authors (X.S.C.) thanks the 
National Natural Science Foundation of China for support under Grant No.
19704005.\\

\newpage
{\bf{Figure Caption}}\\

{\bf{Fig.1.}} Asymptotic $L^{-1} - \xi^{-1}$ plane (schematic plot) 
above $T_c$ for the $\varphi^4$ field-theoretic model at finite 
cutoff $\Lambda$ in the large-$n$ limit in three dimensions where
$L$ is the system size and $\xi$ is the bulk correlation length.
Finite-cutoff effects become non-negligible in the non-scaling
region below the dotted line. This crossover line has a vanishing slope
at the origin and is
determined by Eq. (15) with $\gamma/\nu = 2, \gamma = 2$ and
$a_1 (3,0) = 0.226$ for $d = 3$. Well above this line the cutoff
dependence is negligible in Eq. (11). The arrow indicates an approach
towards bulk critical behavior at constant $0 < t \ll 1$ through the
non-scaling region where Eq. (14) is valid.

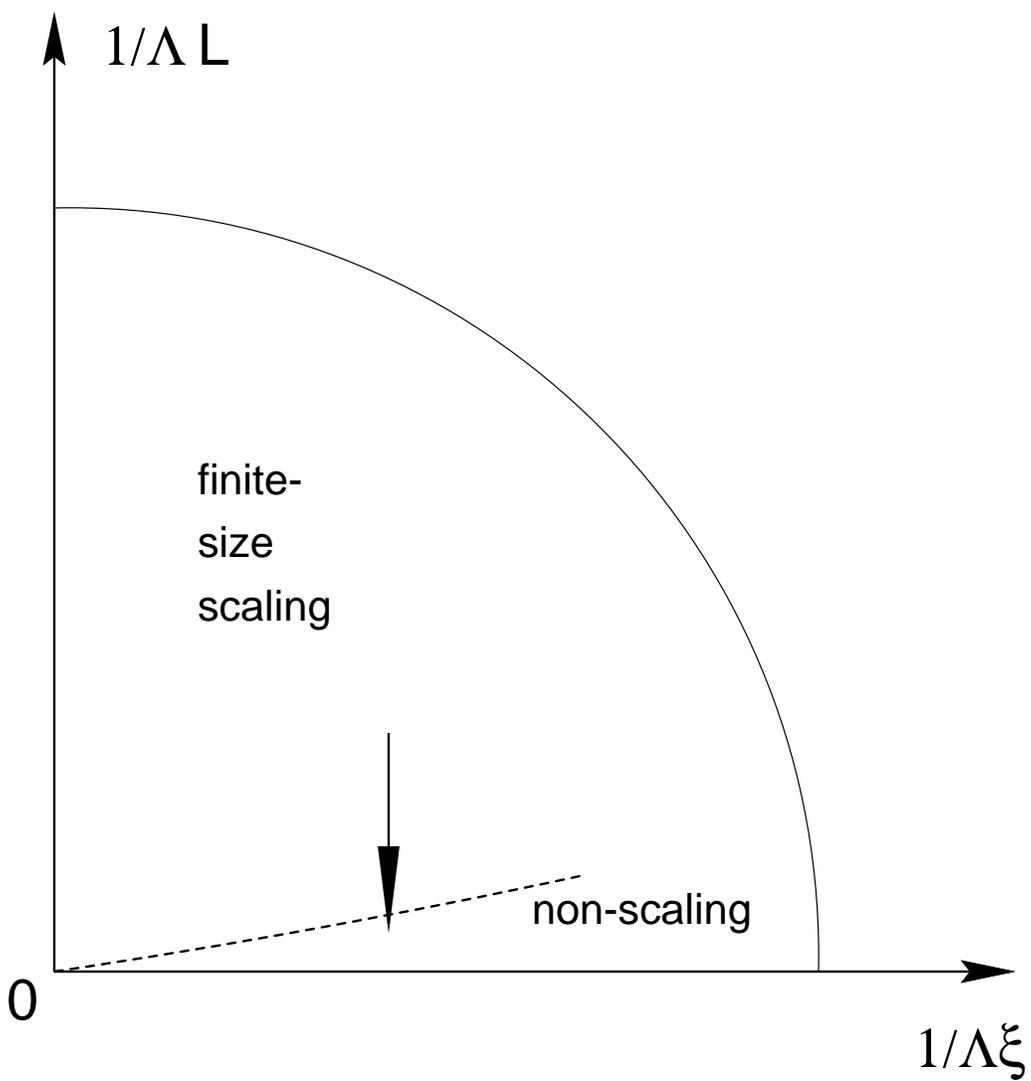
\end{document}